\documentclass[aps,pra,preprint,superscriptaddress,longbibliography]{revtex4-2}

\usepackage{amsmath}
\usepackage{amsthm}
\usepackage{amsfonts}
\usepackage{amssymb}
\usepackage{xcolor,graphicx}
\usepackage{color}
\usepackage[pdftex]{hyperref} 
\usepackage{longtable}
\usepackage{array}
\usepackage{dsfont}

\begin{document}
	\title{Routes to chaos in a class-B laser coupled to a neutral resonator}
	
	\author{L.~E. G\"uemes Frese}
	\email[e-mail: ]{a01154875@tec.mx}
	\affiliation{Tecnologico de Monterrey, Escuela de Ingenier\'ia y Ciencias, Ave. Eugenio Garza Sada 2501, Monterrey, N.L., Mexico, 64849}

	\author{B.~R. Jaramillo-\'Avila}
	\email[e-mail: ]{jaramillo@inaoep.mx}
	\affiliation{CONACYT - Instituto Nacional de Astrof\'isica, \'Optica y Electr\'onica, Calle Luis Enrique Erro No. 1. Sta. Ma. Tonantzintla, Pue. C.P. 72840, Mexico}	

	\author{B.~M. Rodr\'iguez-Lara}
	\email[e-mail: ]{bmlara@tec.mx}
	\affiliation{Tecnologico de Monterrey, Escuela de Ingenier\'ia y Ciencias, Ave. Eugenio Garza Sada 2501, Monterrey, N.L., Mexico, 64849}	
	
	\date{\today}
	
	\begin{abstract}
	We study the dynamics of a class-B semiconductor microring laser coupled to a neutral microring resonator for common fabrication parameters.
	For zero detuning between the resonators, we identify five dynamical regions controlled by the normalized pump rate of the laser and coupling strength between the resonators.
	These regions show stable lasing with phase locking for either sufficiently small or large coupling strength parameter values. 
	Multistability and chaos arise in between these stable regions where at least one fixed point is attractive.
	The transition from stable to unstable lasing regions in the parameter space phase diagram is a crossover.
	\end{abstract}
	
	\maketitle
\section{Introduction}
Arrays of coupled semiconductor microring lasers are of both technological and fundamental importance in optics. 
They may be engineered to emit in synchronization, producing a coherent light source where the power of several microrings is added \cite{Gourley1991,Garcia1999}. 
Coupled in specific configurations, these arrays may harbor topological effects, such as edge states \cite{Bandres2018,Harari2018,Longhi2018,Parto2018,Zhao2018}, or may exhibit dynamic instabilities that lead to chaos. 
While seemingly undesirable, chaos synchronization in these systems enables secure communications, for example, chaos-based cryptography \cite{Argyris2005,Uchida2012,Sciamanna2015,Liu2020a,Liu2020b}.

The most common type of semiconductor laser is the class-B laser \cite{Wieczorek2005,Baili2009,Longhi2018}. 
Its mathematical model arises from Maxwell-Bloch equations in the limit where the lifetime of the active medium polarization is much shorter than the lifetime of both carriers in the media and photons in the cavity \cite{Arecchi1984,Wieczorek2005}. 
A single class-B laser by itself produces a stable field and, therefore, does not lead to dynamic instabilities or chaos. 
When several class-B lasers are coupled, the array can display synchronization \cite{Winful1988,Winful1990}, cluster synchronization \cite{Zhang2019}, chaos and chaos chimeras \cite{Larger2015,Shena2017}. 
These effects are relevant for coherent emission \cite{Gourley1991,Garcia1999}, topological states \cite{Bandres2018,Harari2018,Longhi2018,Parto2018,Zhao2018}, and chaos-based encryption \cite{Liu2020a,Liu2020b}.

Laser arrays are typically built by repeating a unit cell formed by a few coupled semiconductor lasers. 
Understanding the behavior of these unit cells is necessary to describe the larger arrays. 
The dynamical behavior of coupled arrays of identical \cite{Winful1988,Winful1990} and nonidentical \cite{Hohl2017} class-B lasers has been elucidated in the literature. 
For example, a typical nonidentical unit cell consists of a pair of coupled microring lasers pumped at different rates to act in a fashion similar to a gain-loss dimer \cite{Longhi2018,Parto2018,Zhao2018,Padron2020}. 
Here, we focus on a class-B microring semiconductor laser coupled to a passive microring. 
This simple unit cell produces equivalent dynamics to the gain-loss dimer and shows dynamical instabilities and chaos even if a single class-B laser does not. 
The passive microrring has an effect similar to coupling a mirror to the laser, which introduces time-delay effects and leads to dynamical instabilities \cite{Pieroux1994,Li2012}, our model however is free from explicit time-delays. 
In Section \ref{sec:single}, we provide a brief review of the dynamics in a single class-B laser. 
Next, we present our model, study its fixed points and the phase diagram provided by its parameters in Section \ref{sec:fixedpoints}; we focus on the zero detuning case and find two fixed points, one where the phases of the microrings are synchronized and one where they differ by half a cycle. 
In Section \ref{sec:regions}, we characterize the parameter regions where these fixed points are stable, leading to stable synchronization and anti-synchronization. 
We also explore the parameter values where dynamical instabilities and chaos arise. 
Finally, in Section \ref{sec:conclusions}, we close with our conclusions.

\section{Single class-B laser}\label{sec:single}
To provide context, we present a brief review of the dynamics in a single class-B laser \cite{Wieczorek2005,Baili2009},
\begin{align} \label{eq:classB-model}
		i\mathcal{\dot{E}}(t) =&~ \frac{i(1-i\alpha)}{2} \left\{ -\frac{1}{\tau_{p}} + \sigma[n(t)-1] \right\} \mathcal{E}(t),\\
		\dot{n}(t) =&~ R - \frac{n(t)}{\tau_{s}} - \frac{2[n(t)-1]}{\tau_{s}} \vert{\mathcal{E}(t)}\vert^{2},
\end{align} 
in terms of the complex amplitude of the resonant electric field mode $\mathcal{E}(t)$ and the carrier density normalized to transparency $n(t)$. 
The parameters of the model are the linewidth enhancement factor $\alpha$, carrier lifetime $\tau_{s}$, cavity lifetime $\tau_{p}$, differential gain $\sigma$ and normalized pump rate $R$. 
We may safely assume that in any given experimental realization, the carrier lifetime is larger than the cavity lifetime which in turn is larger than the inverse differential gain, $\tau_{s} \gg \tau_{p} \gg \sigma^{-1}$ \cite{Arecchi1984,Baili2009}.
For the sake of simplicity, we move into the frame,
\begin{align} \label{eq:classB-model-A}
	\dot{A}(t) &=~ \left\{ -\frac{1}{\tau_{p}} + \sigma[n(t)-1] \right\} A(t),\\
	\dot{n}(t) &=~ R - \frac{n(t)}{\tau_{s}} - \frac{2[n(t)-1]}{\tau_{s}} A(t),
\end{align}
provided by the squared absolute value of the field mode amplitude $A(t) = \vert \mathcal{E}(t) \vert^{2}$, that allows us to uncouple the dynamics of phase. 
This frame allows us to find two fixed points, 
\begin{alignat}{2}
	& A^{(1)} = 0, \qquad 
	&& n^{(1)} = \tau_{s}R, \\
	& A^{(2)} = \frac{1}{2} \left( \sigma\tau_{p}\tau_{s}R - \sigma\tau_{p} -1 \right), \qquad
	&& n^{(2)} = \frac{1}{\sigma\tau_{p}} + 1.
\end{alignat}
A standard stability analysis of the system requires the eigenvalues of the Jacobian, 
\begin{widetext}
\begin{align}
	\lambda_{\pm} =& \frac{1}{2} \left\{ - \frac{1}{\tau_{p}} + \sigma[n(t)-1] - \frac{1+2A(t)}{\tau_{s}} \right\} 
	\nonumber \\ &
	\pm \frac{1}{2} 
	\Bigg\{ 
	\left[ \frac{1}{\tau_{p}} - \frac{1+2A(t)}{\tau_{s}} \right]^{2} 
	+ \sigma^{2} [n(t)-1]^{2} - 2\sigma[n(t)-1] \left[ \frac{1}{\tau_{p}} - \frac{1-2A(t)}{\tau_{s}} \right] \Bigg\}^{1/2},
	\label{eq:eigenvalues}
\end{align}
\end{widetext}
evaluated at each fixed point to understand the dynamics in the neighborhood of that point \cite{Guckenheimer1983,Murray1989,Jordan2007}. 

The fixed points and Jacobian eigenvalues lead to various critical normalized pump rates. Among these, three are crucial,
\begin{align}
	R^{(A)} &=~ \frac{1}{\tau_{s}} \left( \frac{1}{\sigma\tau_{p}} + 1 \right), \label{eq:first-th} \\
	R^{(B)}_{\pm} &=~ \frac{1}{\tau_{s}} + \frac{2}{\sigma\tau_{p}^{2}} \left[1 \pm \left(1-\frac{\tau_{p}}{\tau_{s}}\right)^{1/2} \right].
		\label{eq:second-th}
\end{align} 
The first one corresponds to the critical normalized pump rate where the absolute value of the field amplitude at the second fixed point becomes non-negative, and therefore physical. 
Above that normalized pump rate, $R^{(A)} < R$, the second fixed point is stable. 
There, for all physical initial conditions the system evolves to stable non-zero values for both the squared absolute value of the field amplitude and a carrier density. 
These values are provided by the second fixed point, $( A^{(2)}, n^{(2)} )$. 
For normalized pump rates below it, $R<R^{(A)}$, neither fixed point has finite and stable field amplitudes; in fact the first fixed point, $(A^{(1)}=0,n^{(1)})$, always has vanishing field amplitude. 
Below this normalized pump rate, for all physical initial conditions the system evolves to zero field amplitude and carrier density value equal to $\tau_{s} R$.
The latter is smaller than the stable carrier density when the laser is pumped above the threshold. 
The other pair of critical normalized pump rates, $R^{(B)}_{\pm}$, arise from evaluating the Jacobian at the second fixed point and finding the normalized pump rates where the Jacobian eigenvalues acquire or lose an imaginary part. 
At these critical normalized pump rates, $R^{(B)}_{\pm}$, the fixed point acquires or loses spiral behavior. 
Under the hierarchy $\tau_{s} \gg \tau_{p} \gg \sigma^{-1}$, these critical rates organize themselves,
\begin{align}
	0 < R^{(A)} < R^{(B)}_{-} < R^{(B)}_{+},
\end{align}
in a manner that highlights four regions leading to a classification of the fixed points, Tab. \ref{tab:singular-points}.
Thus, we observe lasing once the normalized pump rates reach the threshold value $R_{th} = R^{(A)}$.
After this threshold value, the second fixed point $(A^{(2)},n^{(2)})$ behaves like a stable sink in the region $R_{th} < R < R^{(B)}_{-}$, then like a stable spiral sink in the region $R^{(B)}_{-} < R < R^{(B)}_{+}$, and, finally, like a stable sink after that, $ R^{(B)}_{+} < R$ \cite{Wieczorek2005}.

\begin{table*}
\caption{Fixed point classification from the stability analysis of a single class-B laser.}		\label{tab:singular-points}
\begin{ruledtabular}
\begin{tabular}{ccc}
	Region 							& $(A^{(1)},n^{(1)})$			& $(A^{(2)},n^{(2)})$ 			\\ 
	\hline
	$0 < R < R^{(A)}$				& Sink or node (Type I), stable	& Saddle point 					\\ 
	$R^{(A)} < R < R^{(B)}_{-} $	& Saddle point					& Sink or node (Type I), stable \\ 
	$R^{(B)}_{-} < R < R_{+}^{(B)}$	& Saddle point					& Spiral sink, stable	 		\\ 
	$R^{(B)}_{+} < R $				& Saddle point					& Sink or node (Type I), stable
\end{tabular}
\end{ruledtabular}
\end{table*}

\section{Class-B laser coupled to a neutral resonator}\label{sec:fixedpoints}
An usual path to chaos in class-B lasers is to induce feedback by introducing an external mirror \cite{Pieroux1994,Li2012}.
Here, we want to study the dynamics of a class-B laser interacting with a neutral resonator under minimal coupling \cite{Little1997,Liu2005}, 
\begin{align}
	i\mathcal{\dot{E}}_{1}(t) =~& \frac{i(1-i\alpha)}{2} \left\{ - \frac{1}{\tau_{p}} + \sigma[n(t)-1] \right\} \mathcal{E}_{1}(t) 
	+ \omega_{1}\mathcal{E}_{1}+ g\mathcal{E}_{2}(t),\label{eq:E.field-1}\\
	i\mathcal{\dot{E}}_{2}(t) =~& \omega_{2}\mathcal{E}_{2}(t) + g\mathcal{E}_{1}(t),\label{eq:E.field-2}\\
	\dot{n}(t) =~& R - \frac{n(t)}{\tau_{s}} - \frac{2[n(t)-1]}{\tau_{s}} \vert {\mathcal{E}_{1}(t)} \vert^{2}\label{eq:n.field},
\end{align}
where we define the complex amplitudes $\mathcal{E}_{j}(t)$ of the electric field modes localized at each resonator, its corresponding resonant frequency $\omega_{j}$, and the coupling strength between them $g$. 
Moving into a frame rotating at the resonant frequency of the laser,
\begin{align}
	\mathcal{E}_{j}(t) = \sqrt{A_{j}(t)} \exp[i\phi_{j}(t)],
\end{align} 
and rewriting the differential equation set,
\begin{align}
	\dot{A}_{1}(t) =~& \left\{ - \frac{1}{\tau_{p}} + \sigma[n(t)-1] \right\} A_{1}(t) 
	- 2g \sqrt{A_{1}(t) A_{2}(t)} \sin \varphi(t),\\
	\dot{A}_{2}(t) =~& 2g \sqrt{A_{1}(t) A_{2}(t)} \sin \varphi(t),\\
	\dot{\varphi}(t) =~& \Delta\omega - \frac{\alpha}{2}\left\{ - \frac{1}{\tau_{p}} + \sigma[n(t)-1]\right\} 
	+ g \frac{A_{1}(t) - A_{2}(t)}{\sqrt{A_{1}(t) A_{2}(t)}} \cos \varphi(t) ,\\
	\dot{n}(t) =~& R - \frac{n(t)}{\tau_{s}} - \frac{2[n(t)-1]}{\tau_{s}} A_{1}(t),
\end{align}
in terms of the detuning between the resonator frequencies $\Delta \omega = \omega_{2} - \omega_{1}$, the squared amplitudes $A_{j}(t) = \vert \mathcal{E}_{j}(t) \vert^{2}$, and the phase difference $\varphi(t) = \phi_{1}(t) - \phi_{2}(t)$. 
We find four fixed points,
\begin{align}
\left\{ \left( A_{1}^{(j)}, A_{2}^{(j)}, \varphi^{(j)}, n^{(j)} \right) \right\},
\end{align}
with $j = 1, 2, 3, 4$ arising from the combination of the two sign options for the square root in the squared absolute value of the field amplitude inside the neutral resonator $A_{2}^{(j)}$ and the two phase difference options $\varphi^{(j)}$,
\begin{align}
A_{1}^{(j)} =~& \frac{1}{2} \left(\sigma\tau_{p}\tau_{s}R - \sigma\tau_{p} - 1 \right),\\
A_{2}^{(j)} =~& \frac{1}{4g^{2}} \left( 2g^{2} + \Delta\omega^{2} \pm \Delta\omega \sqrt{ 4g^{2}+ \Delta\omega^{2} } \right) 
\left(\sigma\tau_{p}\tau_{s}R - \sigma\tau_{p} - 1 \right), \\
\varphi^{(j)} =~& 0,\pi,\\
n^{(j)} =~& \frac{1}{\sigma\tau_{p}} + 1.
\end{align}
It is straightforward to notice that the case of microrings with identicall resonator frequencies, $\Delta \omega = 0$, reduces to just two fixed points with equal squared absolute value of the field amplitudes, $A_{2}^{(j)}=A_{1}^{(j)}$.
For the sake of simplicity, we focus our analysis on this case.

\begin{figure}[ht]
	\centering
	\includegraphics[width=\columnwidth]{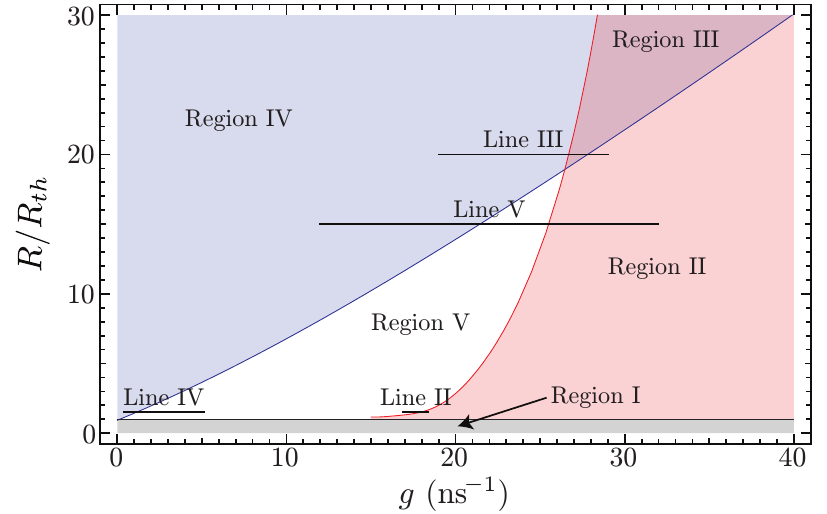} 
	\caption{Fixed point classification from the stability analysis of a class-B laser coupled to a neutral resonator. 
	Region I (gray) is a non-lasing region that lies bellow the threshold normalized pump rate, $R_{th}$. 
	In Region II (red), the first fixed point is a saddle point and the second is a stable sink or stable spiral sink. 
	Region IV (blue) is the opposite. 
	In Region III (overlap), both fixed points are stable sinks or stable spiral sinks.
	In Region V (white), both fixed points are saddle points and we find chaos in it and near its boundaries with Region II and Region IV.}
	\label{fig:Fig1}
\end{figure}

For this zero detuning case, $\Delta \omega = 0$, the squared absolute value of the field amplitudes at either fixed point is negative and therefore unphysical, unless the normalized pump rate is above a threshold value, 
\begin{align}
	R_{th} = \frac{1}{\tau_{s}}\left( \frac{1}{\sigma \tau_{p}} + 1 \right),
\end{align}
identical to that of the single class-B laser. 
This threshold depends only on the gain-resonator parameters and is not influenced by the coupling to the neutral resonator. 
Again, the system will not lase below this threshold normalized pump rate, Region I in Tab. \ref{tab:Tab2} and Fig. \ref{fig:Fig1}(a) where both fixed points are saddle points.
However, the behavior above the lasing threshold is richer. 
We find one stable fixed point in Regions II and IV, and two of them in Region III.
These should provide us with stable lasing.
In Region V, both fixed points behave as saddle points and chaos arises \cite{Guckenheimer1983,Murray1989,Jordan2007}.
In the following, we address with more detail the dynamics of our model in these five regions.

\begin{table*}
\caption{Fixed point classification from the stability analysis of a class-B laser coupled to a neutral resonator.}		\label{tab:Tab2}
\begin{ruledtabular}
\begin{tabular}{ccc}
	Region 		& $(A_{1}^{(1)}, A_{2}^{(1)}, \varphi^{(1)}=0, n^{(1)})$ 	& $(A_{1}^{(2)}, A_{2}^{(2)}, \varphi^{(2)}=\pi, n^{(2)})$	\\
	\hline
	Region I	& Saddle point												& Saddle point 												\\ 
	Region II	& Saddle point												& Sink or spiral sink, stable								\\
	Region III	& Sink or spiral sink, stable								& Sink or spiral sink, stable								\\
	Region IV	& Sink or spiral sink, stable								& Saddle point 												\\
	Region V	& Saddle point												& Saddle point
\end{tabular}
\end{ruledtabular}
\end{table*}

\section{Parameter space phase diagram}\label{sec:regions}
Our dynamical system displays two fixed points in the zero detuning case, $\Delta \omega = 0$. 
We use standard stability analysis \cite{Guckenheimer1983,Murray1989,Jordan2007} to find the parameter values where these fixed points become attractive, pulling nearby trajectories. 
This leads to stable behavior of the system for configurations $\left( A_{1}, A_{2}, \varphi, n \right)$ at and near the fixed point.
To do this, we use typical class-B laser parameters $\alpha = 3$, $\tau_{s} = 4~ \mathrm{ns}$, $\tau_{p} = 40~ \mathrm{ps}$, and $\sigma^{-1} = 1.67~ \mathrm{ps}$ \cite{Longhi2018}, 
and couplings ranging from zero to a few tens of inverse nanoseconds \cite{Zhao2018,Parto2018}. 
Figure \ref{fig:Fig1} displays the regions defined by the dynamical characteristics nearby the two fixed points. 
This parameter space phase diagram depends on 
the normalized pump rate $R$ 
and 
the coupling strength $g$ 
and contains five different regions. 

\subsection{Region I}
The normalized pump rate of the semiconductor laser is below the threshold value $R_{th}$ and no lasing occurs, both fields in the resonators deplete. 
At either fixed point, the squared absolute value of the amplitudes, $A_{1}^{(j)}$ and $A_{2}^{(j)}$, is negative and therefore the fixed points are unphysical. 
The eigenvalues of the Jacobian evaluated on these fixed points predict saddle point dynamics near them. 
This case is similar to a single class-B laser pumped below threshold. 
The system evolves to zero field amplitudes and the carrier density to a value $\tau_{s} R$ smaller than the stable carrier density for a single class-B laser pumped above threshold.

\subsection{Region II}
\begin{figure}[ht]
	\centering
	\includegraphics[width=\columnwidth]{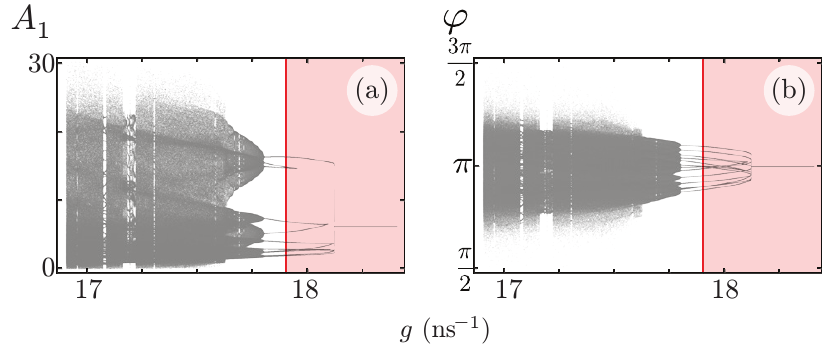} 
	\caption{Bifurcation diagrams for the (a) squared absolute value of the field amplitude $A_{1}$ and (b) phase difference between the field amplitudes $\varphi$ for fixed normalized pump rate and variable coupling strength following Line II in Fig. \ref{fig:Fig1}. The coloring matches that of regions in Fig. \ref{fig:Fig1}.}
	\label{fig:Fig2}
\end{figure}

From this region forward, the two fixed points have identical non-zero amplitude, but their phase difference is zero, $\varphi^{(1)}=0$, at the first fixed point and half a cycle, $\varphi^{(2)}=\pi$, at the second one. 
This region corresponds to large coupling strengths. Here, the first fixed point produces saddle dynamics and the second one produces stable sink dynamics, either spiral or non-spiral depending on the parameter values. 
In other words, we expect stable lasing as the fields reach steady state and their phase difference locks into a value of $\pi$. 
Region II shares a boundary with Region V and Region III on its left. Near these boundaries, we observe 
competition between the attractive component of the saddle point (fixed point 1) and the stable sink (fixed point 2). 
This produces a crossover from stability to multistability and eventually to chaos, Fig. \ref{fig:Fig2}, as the coupling strength $g$ decreases while the normalized pump rate is kept constant, see Line II in Fig. \ref{fig:Fig1}. 

\subsection{Region III}
\begin{figure}[ht]
	\centering
	\includegraphics[width=\columnwidth]{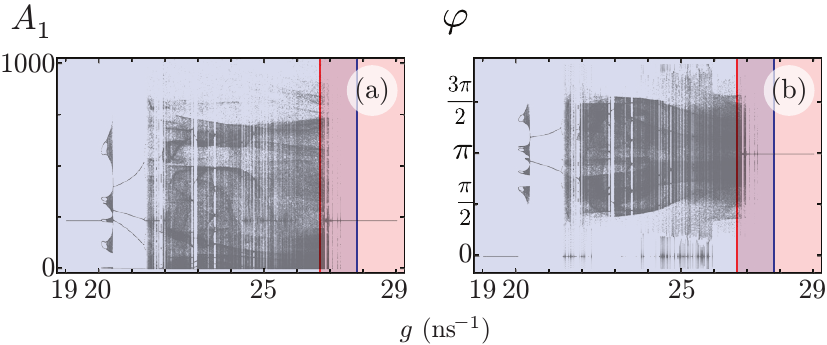} 
	\caption{Same as Fig. \ref{fig:Fig2} but following Line III in Fig. \ref{fig:Fig1}.}
	\label{fig:Fig3}
\end{figure}

Region III appears for large values of the normalized pump rate, $15 R_{th} \lesssim R$. Here, both fixed points have stable behavior either as spiral or non-spiral sinks such that trajectories sufficiently close to either one of them will be attracted and remain there. Region III lies between Regions II and IV where just one of the fixed points is stable. 
We expect equal and stable absolute values for the amplitudes at the resonators, a stable carrier density, and a phase difference locked at either zero or $\pi$, depending on the initial conditions. 
For a fixed normalized pump rate, as the coupling strength $g$ increases, we go from a region on the left where the first fixed point is stable, to a region in the center where both fixed points are stable, to a region on the right where the second fixed point is stable, Line III in Fig. \ref{fig:Fig1}. In both the extreme left and extreme right of this line, we find the expected stable behavior with stable squared absolute value of the field amplitudes and stable phase differences of zero and $\pi$ on the left and right, respectively. However, as we approach the boundaries between Region IV and Region III, we observe multistability and chaos, Fig. \ref{fig:Fig3}, as a consequence of the competition between the attractive parts of the fixed points. 

\subsection{Region IV}
\begin{figure}[ht]
	\centering
	\includegraphics[width=\columnwidth]{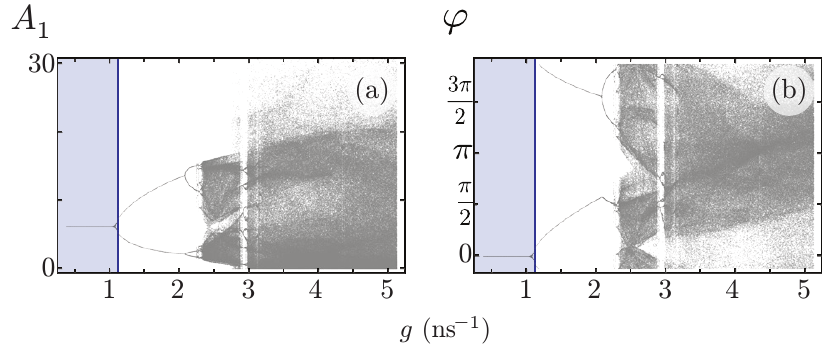} 
	\caption{Same as Fig. \ref{fig:Fig2} but following Line IV in Fig. \ref{fig:Fig1}.}
	\label{fig:Fig4}
\end{figure}

This region corresponds to small coupling strength values. 
The first fixed point produces stable sink dynamics, both spiral or non-spiral depending on the normalized pump rate value, and the second one produces saddle dynamics. 
We expect stable lasing as the fields reach steady state and their phase difference locks to zero. 
Region IV shares a boundary with Region III and Region V on its right. Near these boundaries, we observe competition between the stable sink provided by the first fixed point and the attractive component of the saddle point due to the second fixed point. 
This produces a crossover from stability to multistability and to chaos, Fig. \ref{fig:Fig4}, as the coupling strength $g$ increases with constant normalized pump rate, see Line IV in Fig. \ref{fig:Fig1}.

\subsection{Region V}
\begin{figure}[ht]
	\centering
	\includegraphics[width=\columnwidth]{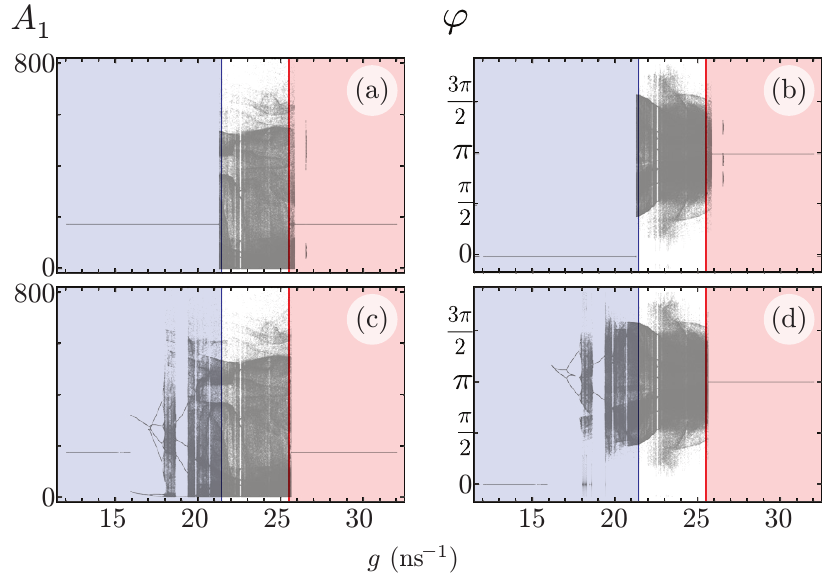} 
	\caption{Same as Fig. \ref{fig:Fig2} but following Line V in Fig. \ref{fig:Fig1}. 
	First (second) row displays initial conditions nearby the fixed point with zero ($\pi$) phase difference.}
	\label{fig:Fig5}
\end{figure}

This region exists for intermediate values of the coupling strength and relatively small normalized pump rates, $R \lesssim 15 R_{th}$. 
Both fixed points present saddle dynamics. 
Competition between the attractive parts of the two saddle points leads to multistability and chaos. 
Region V shares a boundary on the right (left) with Region II (Region IV), where the second (first) fixed point is stable. 
For a fixed normalized pump rate, as the coupling strength $g$ decreases (increases), the stability of the second (first) fixed point in Region II (Region IV) becomes multistable, eventually leading to chaos as we approach Region V, Fig. \ref{fig:Fig1} and Line II (Line IV) in Fig. \ref{fig:Fig2} (Fig. \ref{fig:Fig4}).
Figure \ref{fig:Fig5} displays the full transition from Region IV through Region V and into Region II for initial conditions nearby the first (second) fixed point with zero ($\pi$) phase difference, Fig. \ref{fig:Fig5}(a) and Fig. \ref{fig:Fig5}(b) [Fig. \ref{fig:Fig5}(c) and Fig. \ref{fig:Fig5}(d)].
For a fixed normalized pump rate, as the coupling strength increases, we go from a single attractive first fixed point on the left to a region where neither fixed points is attractive, in the center, to a single attractive second fixed point on the right. 
At the extreme left of this line, we find stable lasing and the phase difference $\varphi$ locks to zero. As we approach the boundary with Region V, multistability and chaos arise. In the center of the line, we observe chaotic behavior. Finally, on the extreme right of the line, we find stable lasing and the phase difference $\varphi$ locks to a value of $\pi$.

We want to stress that the system shows unstable behavior in the parameter regions where at least one fixed point is attractive, Fig. \ref{fig:Fig5}. 
This is to be expected. 
The stability analysis at fixed points is local and only implies that nearby trajectories get pulled to attractive fixed points.
Therefore, whether a trajectory gets pulled or not depends on the initial conditions.

\section{Conclusion}\label{sec:conclusions}
We studied a class-B laser resonator coupled to a neutral one and find a parameter space phase diagram showing a range of dynamical behaviors including phase locked stable lasing, multistability, and chaos. 
This system may serve as a building block for larger arrays of coupled semiconductor lasers; it may be particularly useful for arrays requiring effective gain-loss dimers.

For typical parameters of the class-B laser, we identified five regions in the parameter space phase diagram for zero detuning between the resonators. 
These regions are defined by the characteristics of the two fixed points from the system dynamics and controlled by the values of the normalized pump rate of the laser and coupling strength between resonators.
In the first region, there is no lasing and the resonators are depleted but transient dynamics near the two fixed points shows an unstable phase difference. 
In the second and fourth regions, one fixed point is a stable sink and the other is a saddle point, leading to stable lasing with equal squared absolute value of the field amplitudes and phase difference locked to $\pi$ and zero, in that order, independently of the initial conditions. 
For any given initial condition in the second (fourth) region with large (small) coupling strength values, we are more likely to find stable lasing with phase difference locked to a value of $\pi$ (zero). 
For large normalized pump rate values, there is a region, the third one, with two attractive fixed points, where the phase difference locks to zero or $\pi$ values depending on the initial conditions. 
In the fifth region, we are more likely to find multistable lasing and chaos. 
The transition from stable lasing to chaotic lasing is a crossover occurring in the boundaries between the second and the fifth regions and the fourth and the fifth regions. 

This basic dimer, composed by a standard class-B semiconductor laser coupled to a neutral resonator, shows a rich dynamical landscape where sufficient control of the parameters in the system allows exploring stability with phase locking, multistability, and chaos.

\begin{acknowledgments}
B.~R.~J.~A acknowledges financial support from Catedras CONACYT fellowship program 551.
\end{acknowledgments}

%

\end{document}